\documentclass[aps,jpc,reprint,superscriptaddress, longbibliography]{revtex4-1}

\usepackage{ucs}
\usepackage[utf8x]{inputenc}

\usepackage[english]{babel}
\usepackage{color,graphicx}
\usepackage{amsmath,amsfonts,amssymb,color,fancybox}
\usepackage{units}
\usepackage{epstopdf}

\begin{document}

\title{Statistical mechanics of an elastically pinned membrane: Equilibrium dynamics and power spectrum}

\author{Josip A. Jane\v{s}}
\affiliation{Institut f\"ur Theoretische Physik and Cluster of Excellence: Engineering of Advanced Materials, Friedrich Alexander Universit\"at Erlangen-N\"urnberg, 91052 Erlangen, Germany}
\affiliation{Institut Ru\dj er Bo\v{s}kovi\'c, 10000 Zagreb, Croatia}

\author{Daniel Schmidt}
\affiliation{Institut f\"ur Theoretische Physik and Cluster of Excellence: Engineering of Advanced Materials, Friedrich Alexander Universit\"at Erlangen-N\"urnberg, 91052 Erlangen, Germany}
\affiliation{II. Institut f\"ur Theoretische Physik, Universit\"at Stuttgart, 70569 Stuttgart, Germany}

\author{Udo Seifert}
\affiliation{II. Institut f\"ur Theoretische Physik, Universit\"at Stuttgart, 70569 Stuttgart, Germany}

\author{Ana-Sun\v{c}ana Smith}
\email{author to whom correspondence should be addressed: smith@physik.uni-erlangen.de}
\affiliation{Institut f\"ur Theoretische Physik and Cluster of Excellence: Engineering of Advanced Materials, Friedrich Alexander Universit\"at Erlangen-N\"urnberg, 91052 Erlangen, Germany}
\affiliation{Institut Ru\dj er Bo\v{s}kovi\'c, 10000 Zagreb, Croatia}

\homepage[]{http://puls.physik.fau.de/}

\begin{abstract}
In biological settings membranes typically interact locally with other membranes or the extracellular matrix in the exterior, as well as  with internal cellular structures such as the cytoskeleton. Characterization of the dynamic properties of such interactions presents a difficult task. Significant progress has been achieved through simulations and experiments, yet analytical progress in modelling pinned membranes has been impeded by the complexity of governing equations. Here we circumvent these difficulties by calculating analytically the time-dependent Green's function of the operator governing the dynamics of an elastically pinned membrane in a hydrodynamic surrounding and subject to external forces. This enables us to calculate the equilibrium power spectral density for an overdamped membrane pinned by an elastic, permanently-attached spring subject to thermal excitations. By considering the effects of the finite experimental resolution on the measured spectra, we show that the elasticity of the pinning can be extracted from the experimentally measured spectrum. Membrane fluctuations can thus be used as a tool to probe mechanical properties of the underlying structures. Such a tool may be particularly relevant in the context of cell mechanics, where the elasticity of the membrane's attachment to the cytoskeleton could be measured. 
\end{abstract}
\maketitle

\section{Introduction}

A phospholipid membrane can be easily deformed and exhibits appreciable fluctuations  due to its small elastic constant \cite{helfrich1973,helfrich1978,lipowskysackmann1995, smith2006b,monzel2012,marx2002}. While  occurring on time scales between $10^{-9}$ and $10^{-5}$ s \cite{monzel2016a,fenz2012, betz2012,schmidt2014}, the fluctuations are overdamped by the  surrounding fluid \cite{brochard1975,kramer1971,seifert1993, atzberger2011}. Nonetheless,  mean fluctuation amplitudes of up to $\unit[100]{nm}$ have been observed experimentally \cite{monzel2016b}. In a vicinity of a substrate, these fluctuations are known to contribute to an effective potential which prevents the membrane from non-specifically adhering to the underlying scaffold \cite{helfrich1973, helfrich1984, evans1986, israelachvili1992, raedler1995,lipowskysackmann1995, seifert1997, lorz2007, schmidt2014}. The scaffold in turn affects the hydrodynamic damping of the membrane reflected in changes of the time dependent correlation function and the associated power spectrum, the so-called power spectral density (PSD) \cite{safran2005, betz2009,monzel2015, monzel2016b}. Moreover, in the cellular environment, active processes couple with the membrane fluctuations \cite{prost1998, ramaswamy2000, manneville2001, girard2005, gov2005, lin2006, betz2009, loubet2012, Hanlumyuang2014, alert2015, monzel2015}, resulting in the violation of the fluctuation-dissipation theorem in the activated state of the cell \cite{mizuno2007, ben-isaac2011, turlier2016}. 

Over the last two decades, models for the fluctuations of free membranes, based on the Helfrich energetics and Stokes fluid dynamics, were experimentally confirmed, either by measuring the PSD or its Fourier transform, the time dependent correlation function (for reviews see \cite{monzel2016a,monzel2016b} and references therein). This body of work confirmed the appropriateness of the models based on the Helfrich Hamiltonian \cite{helfrich1973} to capture the equilibrium dynamics of free membranes. However, the presence of the pinning introduces a challenge, which is often circumvented by homogenising the effects of interactions with the scaffold \cite{gov2004b,turlier2016}.
Alternative approaches, where the local pinnings remain explicit, commonly involved simulations \cite{lin2005, reister2011, hu2013, bihr2015}, while the theoretical modelling focused on the static properties of the membrane shape and fluctuations \cite{bruinsma1994, gov2004b, janes2018}. On the other hand, analytic treatments of the membrane-dynamics, even in the context of equilibrium, remained an open problem up to now. First quasi-analytical predictions were obtained for infinitely strong pinnings \cite{lin2004b}, while the more realistic case, where the membrane is attached by proteins which themselves maintain a certain flexibility has not been considered so far. 

In this paper we address this open issue by analytically solving the integro-differential equation governing the motion of a pre-tensed membrane pinned by a single flexible construct. We describe in detail the effect of the pinning on the membrane's equilibrium dynamics. At last, we provide an exact analytical method for calculating the pinning stiffness from the experimentally measured PSD, accounting for the finite resolution of the setup. While constructed in the context of biological membranes, the obtained result can be applied more generally, in the context of bending fluctuations of thin sheets.

\begin{figure*}[tbp]
 \centering
 \includegraphics[width=0.7\linewidth]{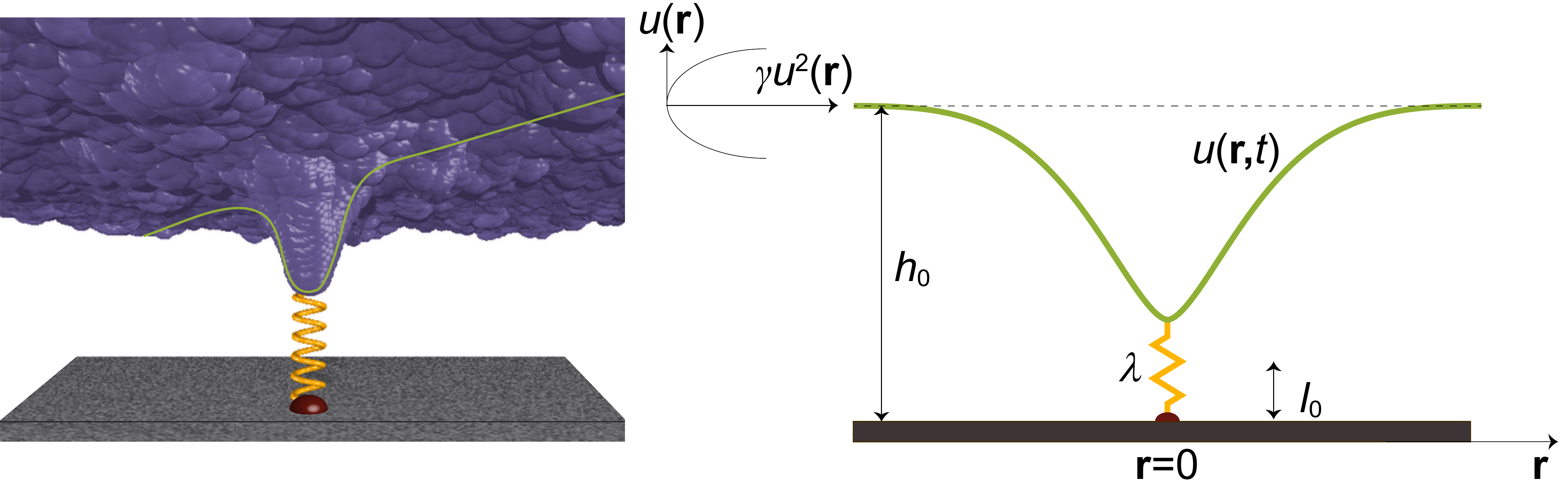}
 \caption{Snapshot from the Langevin simulation of locally pinned membrane fluctuating in a non-specific potential (Left). Sketch of the system (Right). Membrane is residing in a harmonic potential of strength $\gamma$ at $h_0$ separation from a flat substrate and pinned by an elastic spring of stiffness $\lambda$ and rest length $l_0$ positioned at ${\bf r}=0$. 
}
 \label{fig:sketch}
\end{figure*}
%*********************************************************************

\section{Equation of motion}

The system consists of one flexible attachment (harmonic spring of an elastic constant $\lambda$ and rest length $l_0$) that pins a tensed membrane (bending rigidity $\kappa$, tension $\sigma$). The later fluctuates in a harmonic non-specific potential (strength $\gamma$) positioned at the distance $h_0$ above the substrate (Fig. \ref{fig:sketch}). Placing the origin of the coordinate system at the pinning site and into the minimum of the non-specific potential sets the form of the energy functional  \cite{schmidt2012, janes2018} as
\begin{align}
 \mathcal H = \int\limits_A \mathrm{d}{\bf r} &\Bigg[ \frac{\kappa}{2} \left ( \nabla^2u({\bf r},t) \right )^2 + \frac{\sigma}{2} \left ( \nabla u({\bf r},t) \right )^2 + \frac{\gamma}{2} \left ( u({\bf r},t) \right )^2  \nonumber 
 \\
 &+  \frac{1}{2} \lambda\left ( u({\bf r},t) - (l_0-h_0) \right )^2 \delta ({\bf r})\Bigg].
 \label{eq:Translated_Hamiltonian}
\end{align}
Here and throughout the paper, the energy scale $k_\text{B}T$ is set to unity, with Boltzmann constant denoted as $k_\text{B}$ and absolute temperature $T$.

Dynamics of an overdamped membrane in a hydrodynamic surrounding is captured by the Langevin equation \cite{doi1988, seifert1997, granek1997, lin2006b, reister2008, sigurdsson2013, reister2011, bihr2015}
\begin{align}
\frac{\partial u({\bf r},t)}{\partial t} &= \int d{\bf r}' \Lambda({\bf r}-{\bf r}') \left( -\frac{\delta  \mathcal H}{\delta u({\bf r}',t)} + f({\bf r}',t) \right) \nonumber
\\ 
&\equiv \Lambda({\bf r}) * \left( -\frac{\delta  \mathcal H}{\delta u({\bf r},t)} + f({\bf r},t) \right),
\label{eq:Hydrodynamic_equation_for_a_pinned_membrane}
\end{align}
which states that the velocity of the membrane profile $u({\bf r},t)$ is given by a convolution of the hydrodynamic kernel $\Lambda({\bf r})$, the Oseen tensor, with the forces acting on the membrane. External forces on the system are denoted by $f({\bf r},t)$, while the internal forces acting to minimize the Hamiltonian (eq. (\ref{eq:Translated_Hamiltonian})) are given by the first variation \cite{janes2018}
\begin{align}
\frac{\delta  \mathcal H}{\delta u({\bf r},t)} = &\left(\kappa \nabla^4 - \sigma\nabla^2 + \gamma + \lambda \delta ({\bf r})\right) u({\bf r},t) \nonumber \\
- &\lambda (l_0-h_0) \delta ({\bf r}).
\end{align}
Together with eq. (\ref{eq:Hydrodynamic_equation_for_a_pinned_membrane}), this leads to the equation for the dynamics of an overdamped pinned membrane
\begin{align}
\mathcal D \left[u({\bf r},t)\right] = \Lambda({\bf r})* \left[ f({\bf r},t) + \lambda (l_0-h_0) \delta ({\bf r}) \right],
\label{eq:Hydrodynamic_equation_for_a_pinned_membrane_explicit}
\end{align}
with the operator $\mathcal D $ set as
\begin{align}
 \mathcal D = \frac{\partial}{\partial t} + \Lambda({\bf r})*\left(\kappa \nabla^4 - \sigma\nabla^2 + \gamma + \lambda \delta ({\bf r})\right). 
\label{eq:Dynamical_operator}
\end{align}
%***************************************************************************

\section{Time-dependent Green's function}

The solution of eq. (\ref{eq:Hydrodynamic_equation_for_a_pinned_membrane_explicit}) provides the evolution of the membrane profile $u({\bf r},t)$. It is obtained by the integration of forces acting on the membrane, the latter accounted for by the  dynamic Green's function $g({\bf r},t\vert{\bf r}',t')$
\begin{align}
u({\bf r},t) = &\int_{\mathbb{R}^2}\mathrm{d {\bf r}'} \int_{\mathbb{R}}\mathrm{d t}'\  g({\bf r},t\vert{\bf r}',t') \times \nonumber \\
&\times \left( f({\bf r}',t') + \lambda (l_0-h_0) \delta ({\bf r}')\right).
\label{eq:Pinned_membrane_profile_as_aconvolution_of_force_with_the_dynamical_greens_fucntion}
\end{align}
Here the Green's function is defined by
\begin{align}
\mathcal D \left[g({\bf r},t\vert{\bf r}',t')\right] = &\Lambda({\bf r})* \left[ \delta({\bf r}-{\bf r}')\delta(t-t') \right].
\label{eq:Dynamical_Green's_function_of_apinned_membrane_definition}
\end{align}
Besides imposing causality, this equation is subject to homogeneous spatial boundary conditions forcing the membrane in the minimum of the non-specific potential far from the pinning.

\subsection{Free membrane ($\lambda=0$)}

Recognizing spatio-temporal translational invariance of the free membrane system, the corresponding Green's function $g_f({\bf r},t\vert{\bf r}',t') $ can be written in terms of variables $\tilde{t}=t-t'$ and $\tilde{{\bf r}} = {\bf r}-{\bf r}'$ as 
\begin{align}
g_f({\bf r},t\vert{\bf r}',t') = g_f({\bf r}-{\bf r}',t-t')\equiv g_f(\tilde{{\bf r}},\tilde{t}).
\label{eq:Green_spatio_temporal_trans_invariance}
\end{align}
Consequently, for the free membrane eq. (\ref{eq:Dynamical_Green's_function_of_apinned_membrane_definition}) becomes
\begin{align}
& \left[ \frac{\partial}{\partial \tilde{t}} + \Lambda(\tilde{{\bf r}})*\left(\kappa \nabla_{\tilde{{\bf r}}}^4 - \sigma\nabla_{\tilde{{\bf r}}}^2 + \gamma \right) \right] g_f(\tilde{{\bf r}},\tilde{t}) \nonumber \\
&= \Lambda(\tilde{{\bf r}})* \left[ \delta(\tilde{{\bf r}})\delta(\tilde{t}) \right]. 
\label{eq:Trans_invar_free_dynamical_equation}
\end{align}
Fourier transforming eq. (\ref{eq:Trans_invar_free_dynamical_equation}) ($\tilde{{\bf r}}\to {\bf k}$ and $\tilde{t} \to \omega$) upon rearranging yields
\begin{align}
g_f({\bf k},\omega) = \frac{1}{i\omega / \Lambda_k + E_k},
\label{eq:k_omega_Hydrodynamic_equation_for_a_pinned_membrane_greens_fucntion}
\end{align}
where $\Lambda_k$ is the spatial Fourier transform of $\Lambda(\tilde{{\bf r}}$) and
\begin{align}
E_k = \kappa k^4 + \sigma k^2 + \gamma.
\label{eq:eigenvalues}
\end{align}
Finally, transforming back to the spatio-temporal domain (${\bf k}\to \tilde{{\bf r}}$ and $\omega \to \tilde{t}$) provides the spatio-temporal Green's function for the free membrane 
\begin{align}
g_f(\tilde{{\bf r}},\tilde{t})=\int_{\mathbb{R}^2} \frac{d{\bf k}}{(2\pi)^{2}} \int_{\mathbb{R}} \frac{d\omega}{2\pi} \frac{ e^{i{\bf k}\tilde{{\bf r}}}e^{i \omega \tilde{t}}}{i\omega/\Lambda_k + E_k}.
\label{eq:Free_Green_spatio_temporal_trans_invariance}
\end{align}
Integrating over the frequencies gives
\begin{align}
g_f(\tilde{{\bf r}},\tilde{t})&= \int\frac{d{\bf k}}{(2\pi)^{2}} \, \Lambda_k e^{i{\bf k}\tilde{{\bf r}}}e^{-\Lambda_kE_k\tilde{t}} \Theta(\tilde{t})\nonumber
\\
&=\int^{\infty}_0 \frac{dk}{2\pi}\, \Lambda_k k J_0(k\vert\tilde{{\bf r}}\vert) e^{-\Lambda_kE_k\tilde{t}} \Theta(\tilde{t}),
\label{eq:Free_Green_spatio_temporal_trans_invariance_causality}
\end{align}
where $\Theta$ is the Heaviside step function appearing as a consequence of causality. Moreover, $g_f(\tilde{{\bf r}},\tilde{t})$ depends only on the absolute value of $\tilde{{\bf r}}$, as expected.

For $\omega = 0$, the Green's function $g_f(\tilde{{\bf r}},\omega)$ reduces to the static correlation function \cite{janes2018}. Consequently,
\begin{align}
 g_f(\tilde{{\bf r}}=0,\omega=0)  \equiv \frac{1}{\lambda_m} = \frac{\text{arctan}\left(\sqrt{\left(\frac{\lambda_m^{0}}{4\sigma}\right)^{2}-1}\right)}{2\pi\sigma\sqrt{\left(\frac{\lambda_m^{0}}{4\sigma}\right)^{2}-1}}
\label{eq:Free_membrane_fluctuation_amplitude}
\end{align}
represents the fluctuation amplitude, which in the tensionless case reduces to
\begin{align}
 g_f(\tilde{{\bf r}}=0,\omega=0;\sigma=0) \equiv \frac{1}{\lambda^{0}_m} = \frac{1}{8\sqrt{\kappa \gamma}}.
\label{eq:Free_membrane_fluctuation_amplitude}
\end{align}

\subsection{Pinned membrane}

Permanent pinning breaks the spatial, but keeps the temporal translational invariance. Therefore, the Green's function of the pinned membrane must be described by two spatial variables ${\bf r}$ and ${\bf r}'$ and one temporal variable $\tilde{t}=t-t'$
\begin{align}
g({\bf r},t\vert{\bf r}',t') = g({\bf r},t-t'\vert {\bf r}')\equiv g({\bf r},\tilde{t}\vert {\bf r}').
\label{eq:Pinned_Green_temporal_trans_invariance}
\end{align}
In this notation, eq. (\ref{eq:Dynamical_Green's_function_of_apinned_membrane_definition}) for the pinned membrane Green's function becomes 
\begin{align}
\left[ \frac{\partial}{\partial \tilde{t}} + \Lambda({\bf r})*\left(\kappa \nabla^4 - \sigma\nabla^2 + \gamma + \lambda \delta ({\bf r})\right) \right] g({\bf r} ,\tilde{t}\vert {\bf r}' )  \nonumber \\
= \Lambda({\bf r})* \left[ \delta({\bf r}-{\bf r}')\delta(\tilde{t}) \right]. 
\label{eq:Trans_invar_pinned_dynamical_equation}
\end{align}
Fourier transforming (${\bf r}\to {\bf k}$ and $\tilde{t} \to \omega$) and rearranging eq. (\ref{eq:Trans_invar_pinned_dynamical_equation}) gives
\begin{align}
g({\bf k},\omega \vert {\bf r}') = \frac{e^{-i{\bf k}{\bf r}'}}{i\omega / \Lambda_k + E_k} - \lambda g({\bf r}=0,\omega \vert {\bf r}')\frac{1}{i\omega / \Lambda_k + E_k}.
\label{eq:k_omega_Hydrodynamic_equation_for_a_pinned_membrane_greens_fucntion}
\end{align}
Transforming back to the spatial domain (${\bf k}\to {\bf r}$) gives 
\begin{align}
g({\bf r},\omega \vert {\bf r}') = & g_f({\bf r}- {\bf r}',\omega) - \lambda g({\bf r}=0,\omega \vert {\bf r}')g_f({\bf r},\omega).
\label{eq:r_omega_Hydrodynamic_equation_for_a_pinned_membrane_greens_fucntion}
\end{align}
For ${\bf r}=0$ in eq. (\ref{eq:r_omega_Hydrodynamic_equation_for_a_pinned_membrane_greens_fucntion}) we find 
\begin{align}
g({\bf r},\omega \vert {\bf r}') = \frac{g_f({\bf r}',\omega)}{1+\lambda g_f({\bf r}=0,\omega)},
\label{eq:Green_r_omega_pinning_site}
\end{align}
which upon inserting into (\ref{eq:r_omega_Hydrodynamic_equation_for_a_pinned_membrane_greens_fucntion}) yields the Green's function for the pinned membrane in the spatio-frequency domain
\begin{align}
g({\bf r},\omega \vert {\bf r}') = g_f({\bf r}- {\bf r}',\omega) - \lambda \frac{g_f({\bf r},\omega) g_f({\bf r}',\omega)}{1+\lambda g_f({\bf r}=0,\omega)}.
\label{eq:Pinned_Green}
\end{align}
Fourier transforming eq. (\ref{eq:Pinned_Green}) (${\bf r}\to{\bf k}$ and ${\bf r}'\to{\bf k}'$) results in
\begin{align}
g({\bf k},\omega \vert {\bf k}') = (2\pi)^{2} \delta ({\bf k}+{\bf k}') g_f({\bf k},\omega) - \lambda \frac{g_f({\bf k},\omega) g_f({\bf k}',\omega)}{1+\lambda g_f({\bf r}=0,\omega)},
\label{eq:Pinned_Green_k_space}
\end{align}
which is the representation of the Green's function in the Fourier space.

\section{Dynamics of thermal fluctuations}

\subsection{Oseen tensor}

In thermal equilibrium $f({\bf r},t)$ is associated with the stochastic thermal noise characterized by a vanishing mean
\begin{align}
\langle f ({\bf r},t) \rangle =0
\label{eq:Thermal_noise_mean}
\end{align}
and spatio-temporal correlations obeying the fluctuation-dissipation theorem
\begin{align}
\langle f({\bf r},t) f({\bf r}',t^\prime) \rangle = 2\Lambda^{-1}({\bf r}-{\bf r}') \delta (t-t^\prime).
\label{eq:Thermal_noise_variance}
\end{align}
Here $\Lambda^{-1}({\bf r})$ is defined by 
\begin{align}
\Lambda({\bf r})*\Lambda^{-1}({\bf r})=\delta({\bf r}).
\end{align}
To model damping of the membrane due to hydrodynamic interactions with the surrounding fluid close to a wall \cite{doi1988, lin2006b}, we will use the Fourier transform of the Oseen tensor $\Lambda_k$
\begin{align}
\Lambda_k=(4\eta k)^{-1},
\label{eq:Oseen_coefficients}
\end{align}
where $\eta$ is the viscosity of the surrounding fluid. Eq. (\ref{eq:Oseen_coefficients}) is appropriate when the wall is permeable to the fluid. In the presence of an impermeable wall, damping coefficients are modified  \cite{seifert1994a, gov2003}, which in the case of protein mediated adhesion, typically has an effect only on the amplitude of the first few membrane modes \cite{reister2008}.
Furthermore, if the membrane is surrounded by two different fluids with viscosities $\eta_1$ and $\eta_2$, the viscosity $\eta$ in the damping coefficients is replaced by the arithmetic mean  $\eta=(\eta_1 + \eta_2)/2$ \cite{monzel2016b}.

\subsection{Simulation methods}
Eq. (\ref{eq:Hydrodynamic_equation_for_a_pinned_membrane_explicit}) for the membrane dynamics, subject to thermal noise defined with eqs. (\ref{eq:Thermal_noise_mean})-(\ref{eq:Oseen_coefficients}), is the foundation of our Langevin dynamics simulations of the membrane, described previously in full detail \cite{bihr2015}. In the current case, one pinning site is placed in a middle of the simulation box (periodic boundary conditions) of a size of $640\times640$ nm for a tensionless membrane, and a size of $5120 \times  5120$ nm at finite tensions. The simulations are performed with a temporal and lateral resolution of $10^{-9}$ s and $10$ nm, respectively. The membrane height profile is recorded as a function of time and analyzed to extract the membrane shape and correlation functions.

\subsection{Power Spectral Density}

\begin{figure*}[t]
 \centering
 \includegraphics[width=\linewidth]{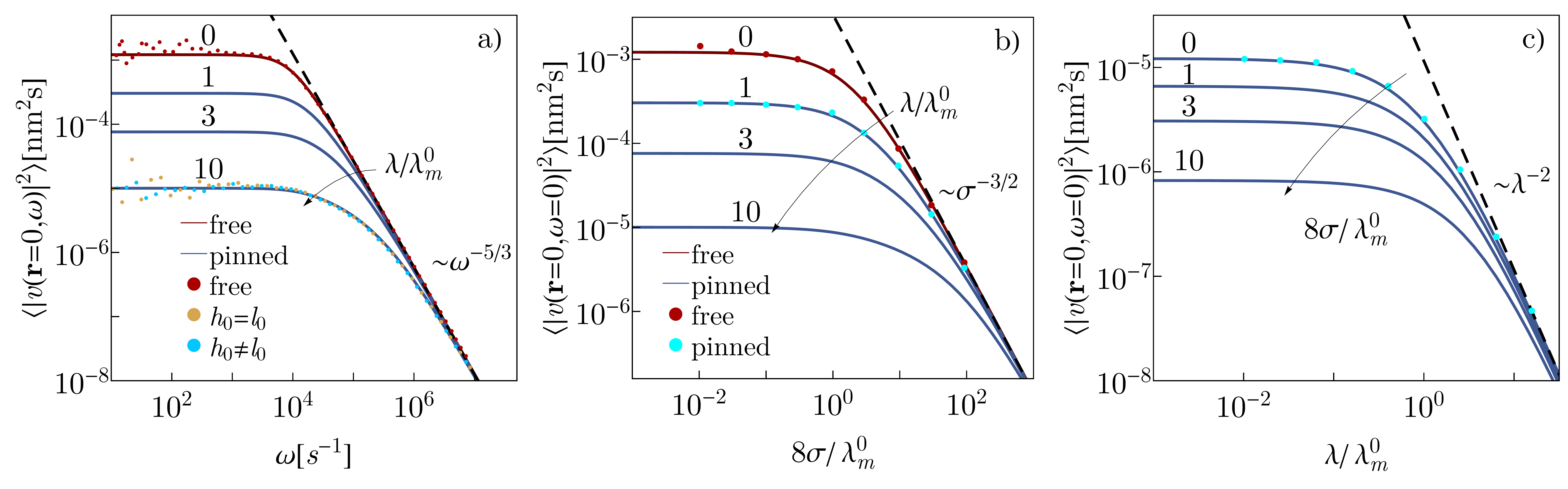}
 \caption{Dynamical properties of a membrane at the pinning site. Comparison of modelling (lines) and simulations (symbols) shows excellent agreement across the entire parameter range. a) Power spectral density of a free membrane (eq. (\ref{eq:PSDu})) (red) and a pinned tensionless membrane eq. (\ref{eq:Single_bond_Power_spectral_density_at_the_pinning}) (blue dashed curves for $\lambda/\lambda_m^0=1,3,10$, increasing in the direction of the arrows). The high-frequency regime of the PSD is unaffected by the pinning and the free-membrane behaviour ($\omega^{-5/3}$) is recovered.
b) Low frequency limit of the PSD (eq. (\ref{eq:Single_bond_Power_spectral_density_at_the_pinning_omega_zero})) as a function of the membrane tension $\sigma$ for different pinning strengths ($\lambda/\lambda_m^0=1,3,10$). For large tensions, a $\sigma^{-3/2}$ dependence is recovered irrespective of $\lambda$.
c) Low frequency limit as a function of the pinning strength for different membrane tensions ($8\sigma/\lambda_m^{0}=1,3,10$).  For large bond stiffness, a $\lambda^{-2}$ dependence is displayed. All curves are plotted for $\kappa=20 k_\text{B}T$, $\gamma=3\times 10^{-7}  k_\text{B}T/$nm$^{4}$ and $\eta=1$ mPas. 
 }
 \label{fig:Exact}
\end{figure*}

We complement the simulations of thermally fluctuating membrane with the analytic calculations based on the Green's function approach (eq. (\ref{eq:Pinned_membrane_profile_as_aconvolution_of_force_with_the_dynamical_greens_fucntion})). We start with   rewriting  eq. (\ref{eq:Pinned_membrane_profile_as_aconvolution_of_force_with_the_dynamical_greens_fucntion}) as
\begin{align}
 u({\bf r},t) =  \langle u({\bf r}) \rangle +  v({\bf r},t),
\label{eq:mean_plus_fluctuations}
\end{align}
where
\begin{align}
v({\bf r},t) = &\int_{\mathbb{R}^2}\mathrm{d {\bf r}'} \int_{\mathbb{R}}\mathrm{d t}'\  g({\bf r},t-t'\vert{\bf r}') f({\bf r}',t')
\label{eq:Fluctuations_Green}
\end{align}
are the fluctuations around the ensemble averaged static profile \cite{janes2018}
\begin{align}
\langle u({\bf r}) \rangle = &\int_{\mathbb{R}^2}\mathrm{d {\bf r}'} \int_{\mathbb{R}}\mathrm{d t}'\  g({\bf r},t-t'\vert{\bf r}')  \lambda (l_0-h_0) \delta ({\bf r}').
\label{eq:Mean_shape_Green}
\end{align}

Transforming ($t\to\omega$) eq. (\ref{eq:Fluctuations_Green}) gives
\begin{align}
v({\bf r},\omega) &= \int_{\mathbb{R}^2}\mathrm{d {\bf r}'}\  g({\bf r},\omega\vert{\bf r}') f({\bf r}',\omega)
\label{eq:Fluctuation_of_omega_space_profile}
\end{align}
from which the PSD $\langle \vert v({\bf r},\omega)\vert ^{2} \rangle$ can be calculated as (see Supplementary Information)
\begin{align}
\langle \vert v({\bf r},\omega) \vert ^{2} \rangle =&\frac{1}{(2\pi)^2}\int_{\mathbb{R}^2}\mathrm{d{\bf k}}\frac{2\Lambda_k^{-1}}{ (\omega/\Lambda_k)^2 + E^2_k} \times \nonumber \\
 & \times \left\vert 1 - \frac{\lambda g_f({\bf r},\omega)}{1+\lambda g_f({\bf r}=0,\omega)} e^{-i{\bf k}{\bf r}} \right\vert ^2.
\label{eq:Power_spectral_density_of_a_membrane_with_a_single_pinning}
\end{align}

With the hydrodynamic coefficients specified as in eq. (\ref{eq:Oseen_coefficients}), eq. (\ref{eq:Power_spectral_density_of_a_membrane_with_a_single_pinning}) becomes
\begin{align}
\langle \vert v({\bf r},\omega) \vert ^{2} \rangle =&\frac{4\eta}{\pi} \int\limits_0^\infty \mathrm d k \frac{k^2}{(4 \eta k \omega)^2 + (\kappa k^4 + \sigma k^2 + \gamma)^2} \times \nonumber \\
 & \times \left\vert 1 - \frac{\lambda g_f({\bf r},\omega)}{1+\lambda g_f({\bf r}=0,\omega)} e^{-i{\bf k}{\bf r}} \right\vert ^2.
\label{eq:Power_spectral_density_of_a_membrane_with_a_single_pinning_hydrodynamic_coeffitients_specified}
\end{align}

In the absence of the pinning ($\lambda =0$), eq. (\ref{eq:Power_spectral_density_of_a_membrane_with_a_single_pinning_hydrodynamic_coeffitients_specified}) becomes homogeneous in space and reduces to the well-known result
\begin{align}
\left\langle \vert v_f(\omega) \vert ^2\right\rangle &= \frac{4\eta}{\pi} \int\limits_0^\infty \mathrm d k \frac{k^2}{(4 \eta k \omega)^2 + (\kappa k^4 + \sigma k^2 + \gamma)^2},
 \label{eq:PSDu}
\end{align} 
which for small and large $\omega$  has the limiting behaviour  \cite{brochard1975, helfer2000, betz2009, betz2012}
\begin{align}
\langle |v_f(\omega)|^2 \rangle =\begin{cases}
	\frac{\eta}{\sqrt{\gamma (\lambda_m^{0}/4 +\sigma)^3 }}, & \omega \ll \omega_0 \\ 
 	\frac{1}{6\sqrt[3]{2\eta^2 \kappa }} \omega^{-5/3}, & \omega \gg \omega_0. 
 	\end{cases}
 \label{eq:PSD_free_limits}
\end{align}
Here,  $\omega_0\equiv \sqrt[4]{\kappa\gamma^3}/\eta$ is a cross-over frequency, defined as the intersection of the lines fitting the low- and high-frequency limits of the spectrum. The low-frequency limit decays with $\sigma^{-3/2}$ for tensions $\sigma \gg \lambda_m^{0}$  (Fig. \ref{fig:Exact}b).

It is clear from Eq. (\ref{eq:Power_spectral_density_of_a_membrane_with_a_single_pinning_hydrodynamic_coeffitients_specified}) that the PSD at the pinning site ${\bf r}=0$  can be recast into
\begin{align}
\langle \vert v({\bf r}=0,\omega)\vert ^{2} \rangle &=\left\vert \frac{1}{1+\lambda g_f({\bf r}=0,\omega)} \right\rvert ^{2} \langle \vert v_f(\omega)\vert ^{2} \rangle.
\label{eq:Single_bond_Power_spectral_density_at_the_pinning}
\end{align}

In agreement with simulations based on eqs. (\ref{eq:Hydrodynamic_equation_for_a_pinned_membrane_explicit}), (\ref{eq:Thermal_noise_mean}) and (\ref{eq:Thermal_noise_variance}) \cite{bihr2015}, eq. (\ref{eq:Single_bond_Power_spectral_density_at_the_pinning}) shows that only the pinning stiffness, and not its length, has an effect on the PSD and that the pinning affects only the low-frequency regime (Fig. \ref{fig:Exact}a). The low-frequency behaviour can be obtained upon combining eq. (\ref{eq:Single_bond_Power_spectral_density_at_the_pinning}) for $\omega=0$ with eq. (\ref{eq:PSD_free_limits}) to yield
\begin{align}
 & \langle \vert v({\bf r}=0,\omega =0)\vert ^{2}  \rangle =  \left( \frac{1}{1+\lambda/\lambda_m}\right) ^{2} \frac{\eta}{\sqrt{\gamma (\lambda_m^{0}/4 +\sigma)^3 }}\nonumber  \\
 = & \begin{cases}
	\left( \frac{1}{1+\lambda/\lambda^{0}_m}\right) ^{2}  \frac{\eta}{\sqrt{\gamma (\lambda^{0}_m/4)^3 }}, &\quad 4\sigma/\lambda_m^{0} \ll 1  \\ 
 	\left( \frac{1}{1+\lambda\ln(\sigma)/(2\pi\sigma)}\right) ^{2}  \frac{\eta}{\sqrt{\gamma \sigma^3 }}, &\quad 4\sigma/\lambda_m^{0} \gg 1.	
 	\end{cases}
\label{eq:Single_bond_Power_spectral_density_at_the_pinning_omega_zero}
\end{align}

Eq. (\ref{eq:Single_bond_Power_spectral_density_at_the_pinning_omega_zero}) shows that the low-frequency spectrum is independent of membrane tension for $ 4\sigma/\lambda_m^{0} \ll 1$ and it decays with $\sigma ^{-3/2}$ for membrane tensions large enough to diminish the effect of the pinning ($\lambda\ln(\sigma)/(2\pi\sigma)\ll 1$) (Fig. \ref{fig:Exact}b). For stiff pinnings ($\lambda/\lambda_m \gg 1 $) the low-frequency limit falls off as $\lambda^{-2}$ (Fig. \ref{fig:Exact}c). On the other hand, for  $\lambda/\lambda_m \ll 1 $, pinning effects vanish even in the low-frequency limit. Interestingly, the cross-over frequency $\tilde{\omega}_0$ for the pinned membrane 
\begin{align}
 \tilde{\omega}_0 & = \left [ 6\sqrt[3]{2\eta^2\kappa}\ \langle \vert v({\bf r}=0,\omega =0)\vert ^{2} \rangle \right ]^{-3/5},
 \label{eq:cross_over_frequency}
\end{align}
defined analogously to the one for the free membrane, becomes sensitive to the elastic properties of the pinning as $\tilde{\omega}_0 \sim \lambda^{6/5}$ and as such increases with the pinning stiffness.

\section{Effect of the finite experimental resolution on the fluctuation spectrum}

In order to compare with experiments, it is necessary to account for the finite temporal and spatial resolutions of the set-up \cite{pecreaux2004,monzel2016b}. Averaging the true membrane profile $ u({\bf r},t) $ over a spatial domain $A$ and a time interval ${\tau}$, gives rise the so-called apparent membrane profile $u^{A}_{\tau} ({\bf r},t) $
\begin{equation}
 u^{A}_{\tau} ({\bf r},t) = \int\limits_0^{\tau} \frac{\mathrm d t^\prime}{{\tau}} \int\limits_A \frac{\mathrm d {\bf r}^\prime}{A} u({\bf r} + {\bf r}^\prime, t+t^\prime),
 \label{eq:h_finres}
\end{equation}
from which it is straightforward to derive the apparent PSD (Supplementary Information, section I. A)
\begin{align}
\langle \vert &v^{A}_{\tau}({\bf r},\omega) \vert ^{2} \rangle = \left(\frac{\sin(\omega {\tau}/2)}{\omega {\tau}/2}\right)^{2}  \frac{2}{(2\pi)^2} \times
\nonumber \\ 
&\times \int_{\mathbb{R}^2}\mathrm{d{\bf k}}\ \frac{\Lambda_k^{-1}}{ (\omega/\Lambda_k)^2 + E^2_k} \left\vert \int\limits_A \frac{\mathrm d {\bf r}'}{A} e^{i{\bf k}({\bf r}+{\bf r}')}\right. \times
\nonumber \\
& \times \left. \left(1 - \frac{\lambda g_f({\bf r}+{\bf r}',\omega)}{1+\lambda g_f(0,\omega)} e^{-i{\bf k}({\bf r}+{\bf r}')}\right) \right\vert ^2.
 \label{eq:apperent_PSD}
\end{align}
The PSD measured around the pinning placed centrally in a circle of radius $R$ is (Supplementary Information, section I. A.1.)
\begin{align}
&\langle \vert v^{R^{2}\pi}_{\tau}({\bf r}=0,\omega)\vert ^{2} \rangle =\left(\frac{\sin(\omega {\tau}/2)}{\omega {\tau}/2}\right)^{2}  \frac{1}{\pi} \times
\nonumber \\
\times &\int\limits^{\infty}_0\mathrm{dk}\ \frac{k \Lambda_k^{-1}}{ (\omega/\Lambda_k)^2 + E^2_k} \rule{0cm}{0.8cm} \frac{4}{R^2}  \nonumber \\ 
\times & \rule{0cm}{0.8cm}  \left\vert  \frac{J_1(kR)}{k} -  \frac{\lambda}{1+\lambda g_f(0,\omega)}\frac{1}{2\pi} \int\limits^{\infty}_0\mathrm{d k'}\frac{J_1(k'R)}{i\omega/\Lambda_{k'}+E_{k'}}  \right\vert ^2.
 \label{eq:apperent_PSD_at_the_pinning}
\end{align}
Eq. (\ref{eq:apperent_PSD_at_the_pinning}) reduces to eq. (\ref{eq:Single_bond_Power_spectral_density_at_the_pinning}) in the limit ${\tau},R\to 0$.

The high frequency regime of the averaged PSD recovers the averaging behaviour of the free membrane - spatial averaging changes the decay from  $\omega^{-5/3}$ to $ \omega^{-2}$ as previously reported \cite{betz2012}, while finite temporal resolution induces an additional attenuation of $ \omega^{-2}$. Hence, the PSD which is subject to both temporal and spatial averaging decays as $\omega^{-4}$.

\begin{figure*}[t]
 \centering
 \includegraphics[width=1 \linewidth]{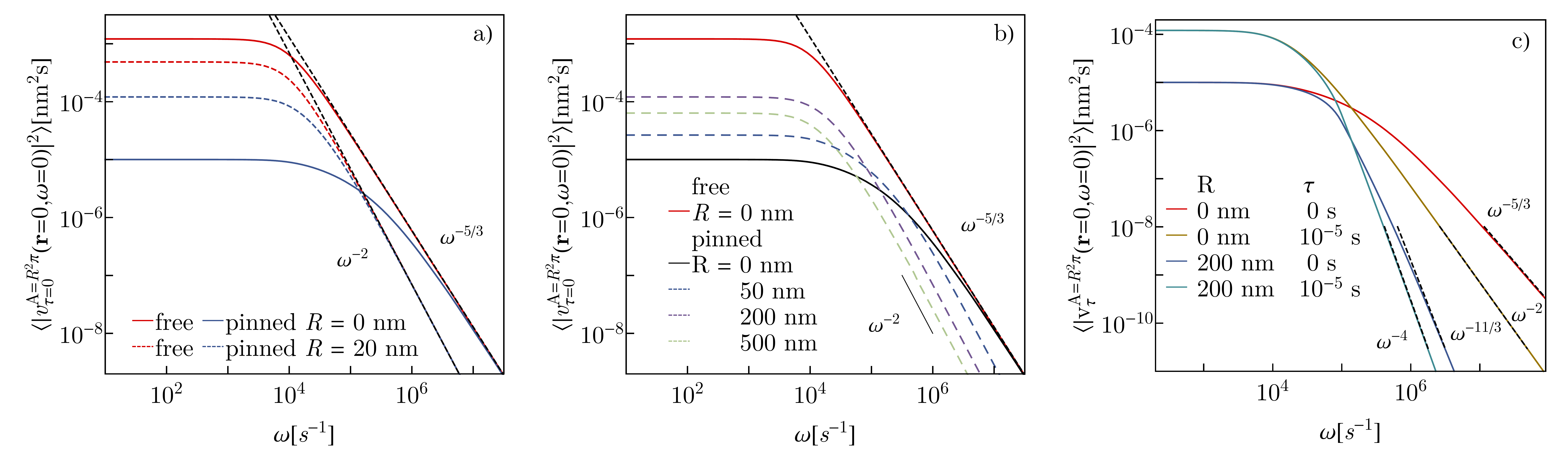}
 \caption{Effect of the finite resolution (averaging over a circle of radius $R$ and a time interval $\tau$) on the PSD at the pinning (eq. (\ref{eq:apperent_PSD_at_the_pinning})). a) Averaging decreases the difference between the free and pinned PSD's, therefore reducing the effect of the pinning on the PSD. Pinning has no effect on the high-frequency part of the spectrum, for both the averaged and the unaveraged spectrum. b) The low-frequency part of the PSD is a non-monotone function of the spatial averaging area. On the other hand, increasing the averaging area always attenuates the high-frequency components, which in this case fall as $\sim \omega^{-2}$ instead of $\sim \omega^{-5/3}$. c) Time averaging introduces oscillatory behaviour of the PSD for which only the envelope of the PSD is shown. The combined effect of the spatial and temporal averaging gives a $\sim \omega^{-4}$ behaviour of the high-frequency regime . Parameters $\kappa=20 k_\text{B}T$, $\sigma = 10^{-20}  k_\text{B}T/\text{nm}^{2}$ and $\gamma=3\times 10^{-7}  k_\text{B}T/$nm$^{4}$, $\lambda /\lambda^{0}_m$=10 and $\eta=\unit[1]{mPas}$.}
 \label{fig:Spatial_averaging}
\end{figure*}

In the low frequency regime, temporal averaging plays no role for $\omega<1/{\tau}$, while the finite spatial resolution has a more complex effect. Due to the interplay with the effects of the pinning, the low frequency amplitude is not a monotone function of the averaging area (Fig. \ref{fig:Spatial_averaging}). Increasing the averaging area up to some critical size (which is approximately the area affected by the pinning) amplifies the low-frequency components, but further increase of the averaging area starts to attenuate them. This can be understood as a competition of two effects; averaging has the effect of attenuating the low frequency components, as can be seen for the free membrane, but at the same time, averaging reduces the effect of the pinning on the PSD, which amplifies the low-frequency components. Obviously, the later effect is stronger up to the critical averaging area size, after which the first effect dominates. Specifically, for $\omega = 0$ we obtain (Supplementary Information, section I.A.1)
\begin{align}
 &\langle \vert v^{R^{2}\pi}_{\tau}({\bf r}=0,0)\vert ^{2} \rangle = \nonumber \\
&= \frac{4\eta}{\pi}\int\limits^{\infty}_0\mathrm{dk}\ \frac{k^2 }{E^2_k} \left\vert  \frac{J_1(kR)}{kR/2} -  \frac{\lambda}{1+\lambda /\lambda_m}s(R) \right\vert ^2,
 \label{eq:single_bond_averaged_PSD_r_zero_omega_zero}
\end{align}
where for brevity purposes we introduce a reduced coefficient $s(R)$
\begin{align}
s(R)&=\frac{1}{R^2\pi\sqrt{\sigma-4\kappa\gamma}} \times  
\nonumber \\
\times &\left(\frac{1-a_-RK_1(a_-R)}{a^2_-}-\frac{1-a_+RK_1(a_+R)}{a^2_+}\right),
 \label{eq:coeff_s_R}
\end{align}
with $a_\pm$
\begin{align}
a_{\pm}&=\left[\frac{\sigma}{2\kappa}\left(1\pm\sqrt{1-\left(\frac{\lambda_m^{0}}{4\sigma}\right)^{2}}\right)\right]^{1/2}.
\label{eq:Free_membrane_Greens_function_coeffitients}
\end{align}

In order to calculate the pinning stiffness $\lambda$ from the PSD, eq. (\ref{eq:single_bond_averaged_PSD_r_zero_omega_zero}) can be inverted,  which upon introduction of coefficients $L$, $a$, $b$ and $c$, yields
\begin{align}
\lambda=\left( \dfrac{1}{L}-\frac{1}{\lambda_m} \right)^{-1},
 \label{eq:lambda}
\end{align}
with
\begin{align}
L=\frac{b}{2a}-\sqrt{\left(\frac{b}{2a}\right)^2-\frac{c}{a}}
 \label{eq:L}
\end{align}
and
\begin{align}
a=&\frac{\pi s^2(R)}{4\sqrt{\gamma(\lambda_m^{0}/4+\sigma)^3}}, \nonumber \\
b=&\frac{4s(R)}{R} \left(\int\limits^{\infty}_0\mathrm{d k}\frac{k J_1(kR)}{E^2_{k}}\right), \nonumber \\
c=&\left(\frac{2}{R}\right)^2\int\limits^{\infty}_0\mathrm{dk}\frac{J_1^2(kR)}{E^2_k} - \frac{\pi}{4\eta} \langle \vert v^{R^{2}\pi}_{\tau}({\bf r}=0,0)\vert ^{2} \rangle .
 \label{eq:coefficients_lambda}
\end{align}
The averaged spectrum is contained in the coefficient $c$. When implemented numerically, eqs. (\ref{eq:lambda}-\ref{eq:coefficients_lambda}) represent a fast and exact method for obtaining information about the pinning stiffness from the experimentally measured PSD. Here we note that the deconvolution of the noise associated with the experimental setting should be performed prior to the extraction of the pinning stiffness. 

\section{Discussion and Conclusion}

%At the pinning site, we can alternatively capture the effect of the pinning  by
%a first-order correction in $\lambda$ to the Oseen tensor
%\begin{align}
% \tilde \Lambda_{q} = \Lambda_{q} \left ( 1+\frac{\lambda}{\lambda_m}\frac{\gamma}{E_q} \right ) .
% \label{eq:Oseen_modified}
%\end{align} 

Our calculation of the Green's function (eq. (\ref{eq:Pinned_Green})) fully resolves the dynamics of an overdamped, permanently pinned membrane in a hydrodynamic surrounding (eq. (\ref{eq:Hydrodynamic_equation_for_a_pinned_membrane})). The solution is general in a sense that it works for any forces acting on the membrane and enables one to study the membrane dynamics in the presence of both non-thermal and thermal perturbations. The later case is resolved in this paper by the calculation of the thermal equilibrium power spectral density (eq. (\ref{eq:Power_spectral_density_of_a_membrane_with_a_single_pinning})). For the specific case of hydrodynamic damping close to a permeable wall, our analytical calculation (eq. (\ref{eq:Power_spectral_density_of_a_membrane_with_a_single_pinning_hydrodynamic_coeffitients_specified})) is verified with Langevin simulations in a broad range of parameters, such as the pinning stiffness, membrane tension, and strength of the non-specific potential, which were allowed to independently vary for several orders of magnitude (Fig. \ref{fig:Exact}). It is shown that the pinning decreases the low-frequency amplitudes of the spectrum and pushes the cross-over frequency (eq. (\ref{eq:cross_over_frequency})) to higher values, while the high-frequency amplitudes remain unaffected (Fig. \ref{fig:Exact}).

Interestingly, the pinned-membrane PSD at the pinning site is given by a product of a free-membrane PSD and a $\lambda$-dependent prefactor (eq. (\ref{eq:Single_bond_Power_spectral_density_at_the_pinning})). Assuming knowledge of the pinning stiffness $\lambda$, this enables inference of the pinned-membrane PSD at the pinning site directly from the free-membrane PSD. This approach has a clear advantage over a direct measurement of the pinned-PSD, as it replaces the pinned-membrane measurement, with a well-established free membrane measurement. On the other hand, if $\lambda$ is not known, it can be easily determined by comparing the pinned- and free-membranes PSDs. The low-frequency limit of the PSD at the pinning site (eq. (\ref{eq:Single_bond_Power_spectral_density_at_the_pinning_omega_zero})) is particularly useful for getting a better understanding of the interplay of the system parameters and shows a $\lambda ^{-2}$ decay. 

These relationships, however, may not be observed experimentally due to the finite resolution of the measurements. Specifically, while the effects of the temporal averaging are simple, significant spatial averaging introduces nontrivial modulations of the spectrum and breaks the relation between the free- and the pinned-membrane PSD given by eq. (\ref{eq:Single_bond_Power_spectral_density_at_the_pinning}). In this regime the pinning stiffness can be inferred from the measured PSD with the use of the spatially averaged spectrum (eqs. (\ref{eq:lambda}-\ref{eq:coefficients_lambda})).

A deep understanding of the mechanics of the pinned membrane is crucial for elucidating the role of more complex pinnings, which under typical biological conditions stochastically bind and unbind from the membrane. The stochasticity of this attachment will have additional effects on fluctuations of the membrane which could not be resolved prior to this investigation, and will be subject to a future study.  The results presented in this paper will thus help establishing the connection between functioning of the protein assembly and the properties of the elastic fluctuating membrane, which is important for understanding of the formation of adhesions. Namely, there is a growing body of evidence that the membrane affects the affinity \cite{huppa2010} and the kinetic rates for protein binding \cite{bihr2015, fenz2017}, which in turn affect the fluctuations and the early stage signalling in developing junctions between cells \cite{perez2008}. 

The model presented here can be used to measure the elasticity of the bond from membrane fluctuations. Hitherto, it was not possible to interpret these measurements accurately, a task that is enabled now by our current work. Such measurements could then be compared to AFM measurements, which are commonly used to study elastic properties of the linkers.

\emph{Acknowledgments:}
A.-S.S and J.A.J. thank ERCStg MembranesAct for support as well as the Croatian Science Foundation research project CompSoLs MolFlex 8238. A.-S.S and D.S. were supported by the Research Training Group 1962 at the Friedrich-Alexander-Universität Erlangen-N\"urnberg.

\bibliography{Literaturliste}

\end{document}